\documentclass[12pt]{article}

\usepackage{graphicx}
\usepackage{color}
\usepackage[colorlinks=true, urlcolor=blue, linkcolor=blue, 
citecolor=blue]{hyperref}

\def \gsim{\mathrel{\vcenter
     {\hbox{$>$}\nointerlineskip\hbox{$\sim$}}}}

\newcommand{\beq}{\begin{equation}}
\newcommand{\eeq}{\end{equation}}
\newcommand{\beqa}{\begin{eqnarray}}
\newcommand{\eeqa}{\end{eqnarray}}
\newcommand{\beqar}{\begin{eqnarray*}}
\newcommand{\eeqar}{\end{eqnarray*}}

\begin{tiny}

\end{tiny} 

\begin{document}
\thispagestyle{empty}

\hfill{\sc UG-FT-314/14}

\vspace*{-2mm}
\hfill{\sc CAFPE-184/14}

\vspace*{-2mm}
\hfill{\sc RM3-TH/14-16}

\vspace{32pt}

\begin{center}

\textbf{\Large A new physics interpretation of the IceCube data}

\vspace{50pt}
Jos\'e Ignacio Illana$^a$, Manuel Masip$^a$, Davide Meloni$^b$
\vspace{16pt}

\textit{$^a$CAFPE and Departamento de F{\'\i}sica Te\'orica y del Cosmos}\\
\textit{Universidad de Granada, E-18071 Granada, Spain}\\
\vspace{10pt}
\textit{$^b$Dipartimento di Matematica e Fisica}\\ 
\textit{Universit\`a di Roma Tre, I-00146 Rome, Italy}\\
\vspace{16pt}

\texttt{jillana@ugr.es, masip@ugr.es, meloni@fis.uniroma3.it}

\end{center}

\vspace{30pt}

\date{\today}

\begin{abstract}
IceCube has recently observed 37 
events of TeV--PeV energies. The angular distribution, 
with a strong preference for downgoing directions, the spectrum,
and the small muon to shower ratio in the data
can {\em not} be accommodated assuming standard interactions 
of atmospheric neutrinos. We obtain an excellent fit, however, 
if a diffuse flux of ultrahigh energy (cosmogenic) neutrinos  
experiences collisions where only a small fraction of the 
energy is transferred to the target nucleon. We show that
consistent models of TeV gravity  or
other non-Wilsonian completions of the standard model provide 
cross sections with these precise features. An increased 
statistics could clearly distinguish our scenario from 
the one assumed by IceCube (a diffuse 
flux of astrophysical neutrinos 
with a $\propto E^{-2}$ spectrum) and establish 
the need for new physics in the interpretation
of the data.

\end{abstract}

\newpage

\section{Introduction}
Neutrinos define the only sector of the standard model (SM)
where some basic
questions have no answer yet. We do not know, for example, 
whether they are Dirac or Majorana spinors, or whether
the sector includes additional sterile modes. 
Although neutrinos are related by the gauge symmetry 
to the electron and the other charged leptons,
the absence of electric charge makes them a very {\em different}
particle. From an experimental point of view their invisibility 
is an obvious challenge that, at the same time, 
provides unexpected opportunities
in the search for new physics. Like protons or photons, neutrinos
are produced with very high energies in astrophysical processes; unlike
these particles, they may cross large distances and reach with 
no energy loss the center of a neutrino telescope like IceCube.
Once there, the relative frequence $\omega_{\rm NP}$
of neutrino interactions involving new physics will be 
enhanced by their small SM cross section:
\beq
\omega_{\rm NP}\approx {\sigma^{\nu N}_{\rm NP}\over \sigma^{\nu N}_{\rm SM}}\,.
\eeq
As we will see, the large target mass in a clean environment 
(only contaminated by atmospheric muons) at telescopes defines 
the ideal ground to probe a class of ultraviolet (UV) completions 
of the SM.

In this article we will be interested in the 37 events
of energy above 30 TeV observed between the years 2010 and 
2013 by IceCube \cite{Aartsen:2013jdh,Aartsen:2014gkd}. 
Their analysis has shown that these events
can {\em not} be explained with standard interactions of atmospheric
neutrinos, even if the lepton flux from charmed hadron decays 
were anomalously high. In the next section we review the IceCube 
analysis and their interpretation, namely, that the origin of
these events is a diffuse flux of cosmic neutrinos with a 
$\propto E^{-2}$ spectrum. We will argue that the data admits other
interpretations, and in Section 3 we describe a new physics 
scenario that does the work. In Section 4 we show that very
{\em soft} collisions of cosmogenic 
neutrinos (with energy around $10^9$~GeV) mediated by this 
new physics would provide an excellent fit to the data, and that
an increased statistics could clearly discriminate this hypothesis
from the standard one.

\section{IceCube data}
The IceCube analysis  isolates neutrino events of energy 
$\gsim 30$~TeV coming from any direction. 
Depending on whether the events include the characteristic track 
of a muon, they are divided into {\em tracks} and {\em showers}.
The directionality in track events is very good, whereas 
the pointlike topology of the showers introduces a $\pm 15^\circ$
uncertainty. 

The analysis tries to eliminate muon tracks 
entering the detector from outside. This also reduces by
a factor of $\approx 0.5$ the number of atmospheric neutrino events 
from downgoing directions. An expected muon background
of $8.4\pm 4.2$ events remains, which seems consistent with the
5 events (one of them containing two coincident 
muons from unrelated air showers) where the  
muon track starts near the detector boundary. We will in 
principle exclude\footnote{We think that these ambiguous events 
could be excluded just by increasing the thresholds in IceTop
and the veto region.} 
events number 3, 8, 18, 28, 32 together with the $8.4\pm 4.2$ 
background from our analysis, assuming that we are then left with 
32 {\em genuine} neutrino interactions inside the IceCube detector,
and we will comment on how the inclusion of these events would affect
our results.

\begin{figure}
\begin{center}
\begin{tabular}{ll}
(a) & (b) \\[-4ex]
\includegraphics{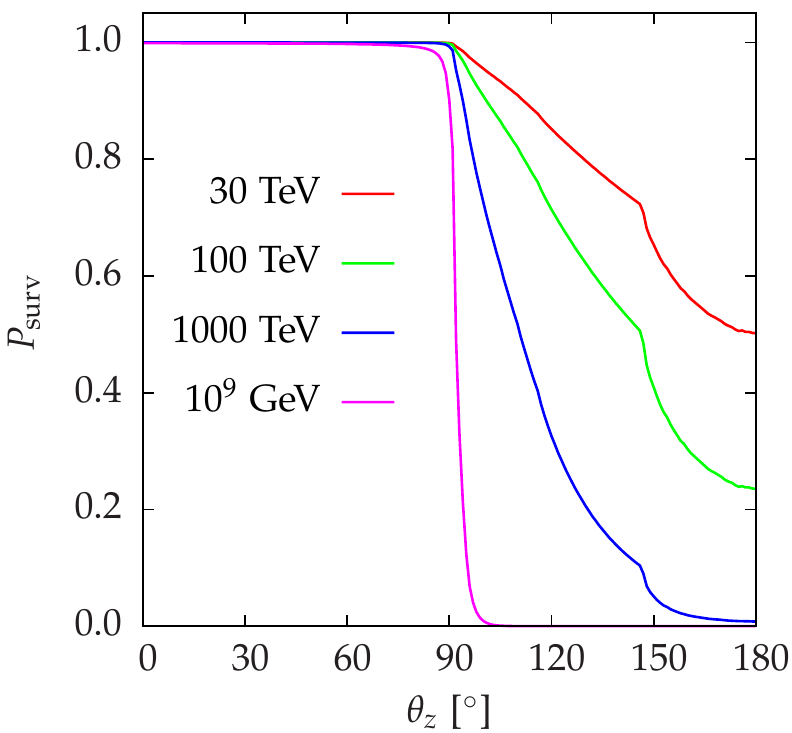} & \includegraphics{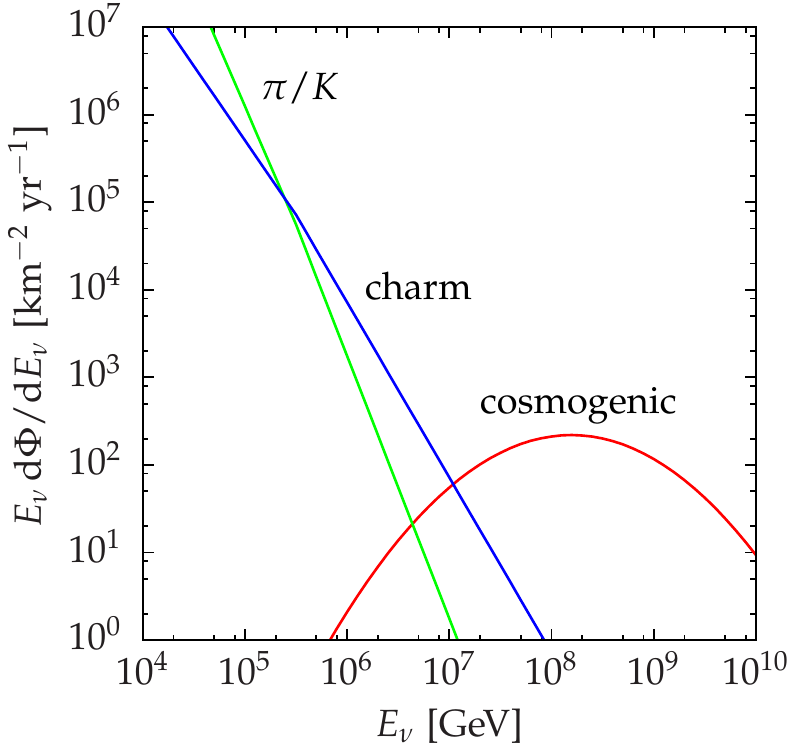}
\end{tabular}
\end{center}
\caption{\it
(a) 
Probability $P_{\rm surv}$ that a neutrino reaches IceCube from
a zenith angle $\theta_z$ for several energies $E_\nu$ (we have used
the $\nu N$ cross section in \cite{Connolly:2011vc}).
(b)
Atmospheric \cite{Illana:2010gh} and cosmogenic \cite{Kotera:2010yn}
neutrino fluxes integrated over all directions and including all flavors.
\label{fig:fig1}
}
\end{figure}

We define two energy bins (30 -- 300~TeV and 300 -- 3000~TeV) 
and three direction bins: {\em downgoing}, which
includes 
declinations $-90^\circ \le \delta < -20^\circ$ ($\delta=\theta_z-90^\circ$),
{\em near-horizontal} ($-20^\circ \le \delta < +20^\circ$) and
{\em upgoing} ($+20^\circ \le \delta < +90^\circ$). 
The Earth is unable to
absorb neutrinos from downgoing and near-horizontal
directions at all the energies of interest, but it becomes opaque
from upgoing directions (see Fig.~\ref{fig:fig1}a), especially in the high
energy bin.
For example, a 100~TeV (1~PeV) neutrino has only 
a 58\% (21\%) probability
to reach IceCube from the $+20^\circ \le \delta < +90^\circ$ bin.

To estimate the number of atmospheric events we will use the 
fluxes in Fig.~\ref{fig:fig1}b. 
We have separated the neutrino flux into the standard component 
from pion and kaon decays plus another component from charmed hadron
decays. The first one has a strong dependence
on the zenith angle (it is larger from horizontal directions)
and is dominated (in an approximate 17:1 ratio) 
by the muon over the electron 
neutrino flavor. The charm component is isotropic and contains
both flavors with the same frequency, together with 
a 2\% $\nu_{\tau}$ component.

\begin{table}
\begin{center}
\scalebox{1.}{
\begin{tabular}{r|c|c|c||c|c|c|c}
          \multicolumn{1}{c}{}
        & \multicolumn{1}{c}{Data} 
        & \multicolumn{1}{c}{$\;$Atm$\;$} 
        & \multicolumn{1}{c}{$\;E^{-2}\;\;$} 
        & \multicolumn{1}{c}{Data} 
        & \multicolumn{1}{c}{$\;$Atm$\;$} 
        & \multicolumn{1}{c}{$\;E^{-2}\;\;$} & \\
\multicolumn{8}{c}{}\\ [-3ex]
\cline{2-7} 
Tracks  &  2 & 0.8 & $\;$0.6$\;$ &  0 & 0.0 & 0.1 & UPGOING \\ [0.5ex]
\cline{2-7}
Showers &  5 & 2.7 & 3.6 &  0 & 0.0 & 0.7 & ($+20^\circ<\delta<+90^\circ$) \\
\cline{2-7}
\multicolumn{8}{c}{}\\[-2ex]
\cline{2-7}
Tracks  &  2 & 3.5 & 1.5 &  0 & 0.0 & 0.5 & NEAR HORIZONTAL \\
\cline{2-7}
Showers &  8 & 5.9 & 6.4 &  1 & 0.2 & 2.6 & ($-20^\circ<\delta<+20^\circ$) \\
\cline{2-7}
\multicolumn{8}{c}{}\\[-2ex]
\cline{2-7}
Tracks  &  0 & 0.2 & 1.6 &  0 & 0.0 & 0.6 & DOWNGOING \\
\cline{2-7}
Showers & 11 & 0.6 & 6.5 &  3 & 0.0 & 2.9 & ($-90^\circ<\delta<-20^\circ$) \\
\cline{2-7}
\multicolumn{8}{c}{}\\[-2.5ex]
 \multicolumn{1}{c}{}
&\multicolumn{3}{c}{30 -- 300~TeV} & \multicolumn{3}{c}{300 -- 3000~TeV}
\end{tabular}
}
\end{center}
\vspace{-0.5cm}
\caption{\it Data, atmospheric background, and best fit of the excess 
with a $E^{-2}$ diffuse flux at IceCube in 988 days.
\label{tab:events1}
}
\end{table}

The 32 neutrino events and our estimate for the atmospheric 
background can be found in Table~\ref{tab:events1}. An inspection of the data
reveals two clear features:

\begin{enumerate}
\item The number and distribution of tracks
is well explained by atmospheric neutrinos. In the low-energy
bin there are 4 tracks
from upgoing and near-horizontal directions for an expected
background of 4.3, whereas at higher energies there are
no events but just 0.06 tracks expected. If we added the 5
downgoing tracks excluded in our analysis together with the
$8.4\pm 4.2$ muon background, we would expect a total of 12.9 track
events and find only 9 in the data: again, no need for extra tracks.
\item There is an excess of showers that is
especially significant from downgoing directions. At low 
energies we find 11 events for 0.6 expected, and in the 
300 -- 3000~TeV bin there are 3 showers for a 0.04 
background. If we include near-horizontal directions we
obtain a total of 23 events for just 6.7 expected.
\end{enumerate}

IceCube then proposes a fit to the excess 
using a diffuse flux of 
astrophysical neutrinos with spectrum proportional to $E^{-2}$
(also in Table~\ref{tab:events1}). We find
that this $E^{-2}$ hypothesis has two generic implications. First,
it gives around 4.5 showers per track. Second, it implies
a very similar number of downgoing and near-horizontal
events (see Table~\ref{tab:events1}). 
To compare it with the data we just
subtract the atmospheric background. We
obtain:

\begin{itemize}
\item 
An excess of 18.6 showers
(28 observed, 9.4 expected) while no tracks (4 observed, 4.5
expected). The IceCube hypothesis introduces 18.4 showers 
and 4.2 tracks.
\item  An excess of 13.2 downgoing events 
but just 1.4 extra events from near-horizontal directions.
The $E^{-2}$ diffuse flux proposed by  IceCube predicts,
respectively, 11.6 and 11.0 events.
\end{itemize}
Therefore, although the statistical significance of these 
deviations is not conclusive yet \cite{Chen:2013dza}, it is apparent that
other possibilities may give a better fit.
In particular, we will define a new physics scenario
that only introduces near-horizontal and downgoing showers
(in a 1:2 ratio) with no new muon tracks from any directions.

\section{A consistent model of TeV gravity}
Consider a model of gravity \cite{ArkaniHamed:1998rs} 
with one flat extra dimension $y$
of radius $R$ and a fundamental 
scale\footnote{We will follow the notation in \cite{Giudice:2001ce}:
$\bar M_P= M_P /\sqrt{8\pi}$ and $\bar M_D=  M_D / (2\pi)^{n/(2+n)}$, 
where $n=D-4$ is the number of extra dimensions. Using this
notation $G_D=G_N(2\pi R)^n =1/(8\pi \bar M_D^{2+n})$ for any 
value of $n$, including $n=0$.} 
$\bar M_5\approx 1$~TeV. Since the generalized Newton's 
constant in $D$ dimensions is $G_D=V_nG_N$, this setup 
requires a very large extra dimension: 
\beq
V_1=2\pi R= {\bar M_P^2\over \bar M_5^3}
\eeq
{\em i.e.}, $R \approx (10^{-27}\;{\rm GeV})^{-1}\approx 1$ AU. 
A change from $1/r$ to $1/r^2$ in the gravitational potential
at such large distances would of course have been observed. 
The model is also excluded by astrophysical \cite{Hannestad:2003yd}
and cosmological \cite{Hannestad:2001nq}
bounds. This can be understood in terms of the 
Kaluza-Klein (KK) modes, of mass $m_n=nm_c$ with $m_c=1/R$.
Although each excitation couples 
very weakly ($\propto \bar M_P^{-1}$) to matter, the large multiplicity
of light states during primordial nucleosynthesis  or 
supernova explosions 
would introduce unacceptable changes in the dynamics.

We intend to solve these problems while keeping the main features
of the model. In particular,
\begin{enumerate}
\item 
We will keep the same 
$\bar M_5\approx 1$~TeV. $M_5$ is the scale ($\mu$) 
where gravity becomes strong: the number of light KK 
modes ($2 \mu/m_c$) times their coupling squared 
to matter ($\mu^2/\bar M_P^2$) gives an amplitude of
order 1 at $\mu=M_5$.
\item
In order to avoid astrophysical and cosmological bounds,
we will increase the mass of the first KK mode and the
mass gap between excitations from $m_c=1/R$
to $m_c\ge 50$~MeV. 
Obviously, since now there are {\em less} KK gravitons,
consistency with the previous point will require that
the coupling squared of each mode is increased by a factor of 
$m_c \bar M_P^2/M_5^3$. 
\end{enumerate}
Notice that doing that the gravitational potential at
distances $r<1/m_c$ (approximately 4~fm for
$m_c=50$~MeV) will be exactly the same as in the case
of one very large compact dimension: the smaller 
density of KK modes is exactly compensated by their larger 
coupling. The main difference is that now gravity 
becomes 4-dimensional at distances much shorter than
before, $r>1/m_c$ instead of $r>1$ AU. 

The framework just outlined would be an explicit realization
of the UV completion by classicalization discussed
in \cite{Dvali:2010jz,Dvali:2012mx}, and it has been defined 
by Giudice, Plehn and Strumia in \cite{Giudice:2004mg} as follows
(see also \cite{Borunda:2009wd}). Let us deform the 
flat circle described above to an orbifold by identifying
$y\to -y$, and let us place 4-dim
branes at $y=0$ (IR brane) and $y=\pi R$ (UV brane). We will
also introduce a (slight) warping along the 
extra dimension:
\beq 
{\rm d} s^2 = e^{2\sigma(y)} \eta_{\mu\nu}\, {\rm d}x^\mu {\rm d}x^{\nu}
+ {\rm d} y^2\,,\;\;\; \sigma(y)\equiv k \, |y|\,.
\eeq
The 4-dim Planck mass is then given by
\beq 
\bar M_P^2 = {\bar M_5^3\over k} \left( e^{2 k \pi R} -1\right)\,.
\label{warp}
\eeq
If the 5-dim curvature is $k\ll 1/R$ we recover in Eq.~(\ref{warp}) 
the flat case, 
\beq 
\bar M_P^2 \approx \bar M_5^3 \, 2 \pi R\,,
\eeq
together with 
a tower of KK gravitons with mass $m_n=n/R$ and 
coupling\footnote{Notice that the orbifolding 
projects out half of the KK modes but also increases by a factor of
$\sqrt{2}$ their coupling to matter.} $\approx
\sqrt{2} \mu/\bar M_P$. We will take, however, the
opposite limit: $k$ larger than $R^{-1}$ but still much smaller than
$\bar M_5$. For example, we obtain $\bar M_5=1$ TeV in Eq.~(\ref{warp}) 
for $k=50$ MeV and $R=(5$ MeV$)^{-1}=40$ fm.
The curvature has then two main effects on the 
KK gravitons \cite{Giudice:2004mg}: their masses 
become proportional to $\pi k\equiv m_c$,
\beq
m_n\approx \left( n+{1\over 4} \right) k \pi 
= \left( n+{1\over 4} \right) m_c\,,
\eeq
and their 5-dim wave function is {\em pushed} towards the
IR brane. 
Assuming that quarks and neutrinos are located there,
this will translate into a larger coupling of all the gravitons
to matter, 
\beq
{\sqrt{2} \over \bar M_P}\to  \sqrt{k\over \bar M_5^3}
\approx  \sqrt{2 m_c\over M_5^3}\,.
\eeq
This is exactly 
the factor discussed above. In short, this
TeV gravity model has just one extra dimension, a low fundamental  
scale $M_5\approx 1$~TeV, and an arbitrary mass $m_c\ge 50$~MeV 
for the first KK mode. Given the (approximately) constant 
mass gap between resonances and their enhanced coupling to
matter, the model gives at distances $r<m_c^{-1}$ 
the same gravitational potential as a model with one flat extra
dimension of length $L\approx 1$ AU, 
while at $r>m_c^{-1}$ it implies Newton's 4-dim gravity.

Once the setup has been justified, we can consider
graviton-mediated collisions
at center of mass energies $s > M_5^2$, {\em i.e.}, in the
transplanckian regime \cite{Emparan:2001kf}. 
In particular, we will be interested 
in scatterings with large impact parameter:  distances longer 
than the typical ones to form a
black hole (and thus with a larger cross section) but 
still shorter than $1/m_c$, 
so that gravity is still purely 5-dimensional. In these processes
the incident neutrino  interacts with a 
parton in the target nucleon, 
transfers a small fraction\footnote{We use the same symbol
$y$ for the inelasticity and the label of the extra dimension
hoping that it does not mislead the reader.} 
$y=(E_\nu-E'_\nu)/E_\nu$
of its energy and keeps going with almost the same energy.
Using the eikonal approximation the amplitude for this process 
can be calculated in impact parameter space
as a sum of ladder and cross ladder diagrams. It turns out 
that \cite{Giudice:2001ce,Illana:2005pu}
\beq
{\cal A}_{\rm eik}(\hat s,q)=4\pi \hat s b_c^2\; F_1(b_c q)\;,
\label{eikonal}
\eeq
where $\hat s$ and $\hat t$ refer to the Mandelstam
variables at the parton level, $y=-t/s$, $q=\sqrt{-\hat t}$,
$b_c= \hat s/(4\,M^3_5)$ and 
\beq
F_n(u)=-i
\int_0^\infty {\rm d}v\;v\; J_0(uv)
\left( e^{iv^{-n}} -1 \right)\;.
\label{f5}
\eeq
\begin{figure}
\begin{center}
\begin{tabular}{ll}
(a) & (b) \\[-4ex]
\includegraphics{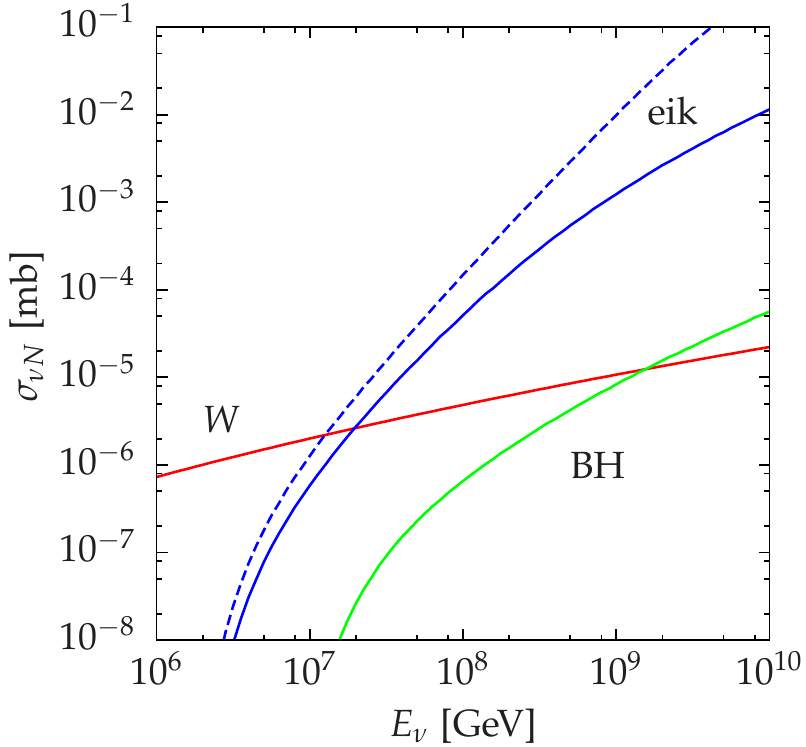} & \includegraphics{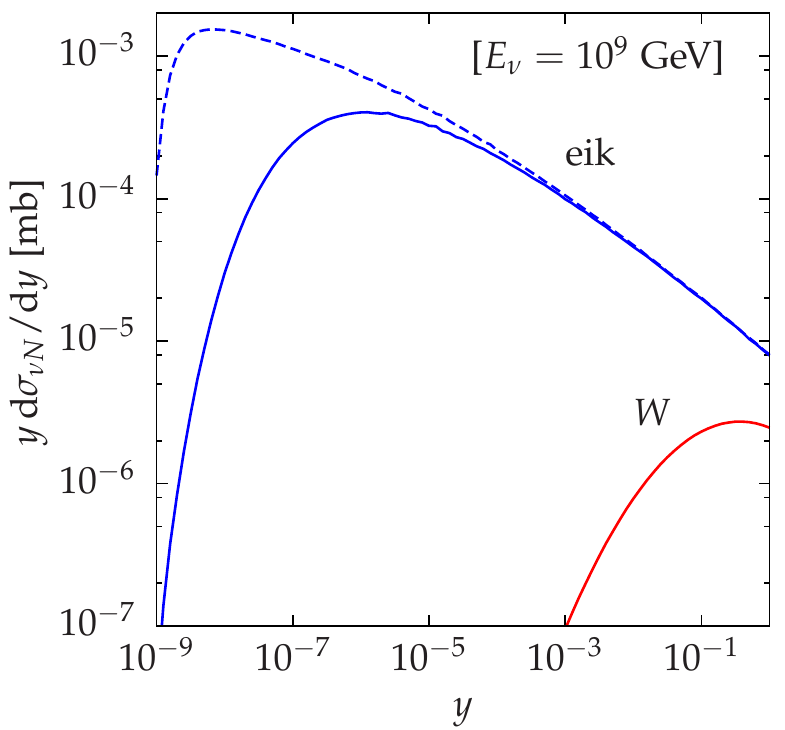}
\end{tabular}
\end{center}
\vspace{-0.5cm}
\caption{\it 
(a) 
$\nu N$ cross sections for proceses mediated
by TeV-gravity  and by $W$ exchange. 
(b)
Differential cross sections 
$y\, {\rm d} \sigma/{\rm d} y$ for $E_\nu=10^9$~GeV.
In both panels $M_5=1.7$~TeV and $m_c=5$~GeV (solid), 50~MeV (dashed).
\label{fig:fig2}
}
\end{figure}
The differential $\nu N$ cross section that we propose 
is then
\beq
\frac{{\rm d}\sigma^{\nu N}_{\rm eik}}{{\rm d}y}=\int^1_{M^2_5/s} \!\! {\rm d}x\ xs\ 
\pi b^4_c \left|F_1(b_cq)\right|^2 \ e^{-2m_c/\mu}
\sum_{i=q,\bar{q},g}f_i(x,\mu)\,,
\eeq
where $\mu=1/b_c$ if $q<b_c^{-1}$ or 
$\mu=\sqrt{q/b_c}$ otherwise 
is the typical inverse distance in the collision and
we have included a Yukawa suppression at distances larger than $m_c$
(a numerical fit gives 
$\left|F_1(u)\right|^2 = 1/ (1.57 u^3 + u^2)$).
Fig.~\ref{fig:fig2} summarizes its main features. At low energies 
the new physics is negligible, and neutrinos interact with matter
only through $W$ and $Z$ exchange. Above an energy threshold 
$E_\nu=M_5^2/(2m_N)\approx 10^6$~GeV the gravitational 
cross section grows fast, and it becomes  much larger than 
the standard one at $E_\nu\approx 10^8$~GeV. 
This large cross section, however, is very soft 
(see Fig.~\ref{fig:fig2}b): the neutrino mean free path in
ice becomes short ($\approx 10$ km at $10^9$~GeV), but the 
fraction of energy deposited in each interaction is
small ($\langle y \rangle \approx 10^{-5}$). Notice that, in addition to
$W$-mediated collisions, only the short distance 
interactions of $\langle y\rangle\approx 1$ or those resulting into
a mini-black hole (see our estimate in Fig.~\ref{fig:fig2}a)
are able to {\em stop} the neutrino when it propagates 
through matter.

The low-$y$ end of the differential cross section 
in Fig.~\ref{fig:fig2}b is regulated
by the arbitrary parameter $m_c$. If the mass of the lightest
KK graviton is around $50$~MeV, then a $10^{10}$~GeV neutrino
would have several TeV energy depositions inside a km of ice,
whereas values $m_c\approx 5$~GeV prevent the total 
cross section from reaching very large values.

Let us finally mention that, although our framework is unconstrained
by astrophysics and cosmology, collider bounds would 
be similar to the ones obtained in any TeV gravity model. If a black 
hole is created one expects a high-multiplicity event with some jets
and leptons of large $E_T$. These bounds, however, are weak for
a low number of extra dimensions \cite{Aad:2014gka} (just one in 
our case) and very model dependent (for example, on the angular momentum 
of the black hole or on the minimum mass --in units of $M_D$-- that
it should have). Other experimental 
constraints could be obtained from the missing $E_T$ 
associated to the production of the massive gravitons. In particular,
an analysis of LEP data suggests bounds between 1.5 
and 2.4 TeV on $M_5$ \cite{Giudice:2004mg}. These bounds are also quite
model dependent: they become weaker if, for example, we
{\it hide} the right-handed electron in the UV brane. Notice that
the particular model with just one extra dimension under study
has not attracted the interest of the experimentalists at the LHC, as
they may consider it excluded by astrophysical observations.
In any case, we would like to emphasize that the very soft 
collisions that we propose, with the incident particle 
losing a very small fraction of energy, are invisible in colliders:
they imply new ultraforward physics, at rapidities out of reach there. 
The ideal place to test such type of cross sections is not colliders,
it is IceCube.

\section{Fit of the IceCube data}
To fit the IceCube data we will use the cosmogenic 
neutrino flux in Fig.~\ref{fig:fig1}b \cite{Kotera:2010yn} and the
eikonal collisions discussed in the previous section.
The cosmogenic flux is mostly produced in collisions
of cosmic rays with the CMB radiation, and it
consists of a few hundred
neutrinos of energy between $10^8$ and $10^{10}$~GeV per km$^2$ 
and year. 

Cosmogenic neutrinos can reach the center of IceCube from
zenith angles $\theta_z\le 90^\circ$ and deposit there a small fraction
of energy through an eikonal scattering. Notice that these
soft collisions do not {\em destroy} the incident 
neutrino, which could actually interact once or several times 
in the ice before reaching the detector. However, 
short distance (both standard and gravitational) interactions
will always prevent cosmogenic neutrinos from reaching IceCube 
from high inclinations ({\em i.e.}, upgoing directions). 
For example, a $10^9$~GeV neutrino has a cross section 
$\sigma_{\nu N}^{CC}\approx 10$ nb for $W$ exchange 
with a nucleon, or $\sigma_{\nu N}^{BH}\approx 8$ nb
to produce a black hole\footnote{We take $M_5=1.7$~TeV and a
geometrical cross section to produce a mini-black hole.} 
through short distance gravitational interactions. However, 
the cross section for an eikonal interactions is much larger,
$\sigma_{\nu N}^{eik}\approx 1$ $\mu$b.
Therefore, soft (long-distance)
gravitational collisions would introduce in IceCube an excess 
of downgoing and near-horizontal showers only.

In Table~\ref{tab:events2} we give the number of eikonal events for the diffuse
cosmogenic flux in Fig.~1 that corresponds to
$M_5=1.7$~TeV, $m_c=1$~GeV in a 3 year period. 
For comparison, we include our estimate 
using the diffuse $E^{-2}$ flux proposed
by IceCube.

\begin{table}
\begin{center}
\begin{tabular}{r|c|c|c|c||c|c|c|c|c}
          \multicolumn{1}{c}{}
        & \multicolumn{1}{c}{Data} 
        & \multicolumn{1}{c}{$\;$Atm$\;$} 
        & \multicolumn{1}{c}{$\;E^{-2}\;$} 
        & \multicolumn{1}{c}{$\;\;$NP$\;\;$} 
        & \multicolumn{1}{c}{Data} 
        & \multicolumn{1}{c}{$\;$Atm$\;$} 
        & \multicolumn{1}{c}{$\;E^{-2}\;$} 
        & \multicolumn{1}{c}{$\;\;$NP$\;\;$} & \\
\multicolumn{10}{c}{}\\[-3ex]
\cline{2-9}
Tracks  &  2 & 0.8 & 0.6 & 0.0 &  0 & 0.0 & 0.1 & 0.0 & UPGOING \\
\cline{2-9}
Showers &  5 & 2.7 & 3.6 & 0.0 &  0 & 0.0 & 0.7 & 0.0 & 
($+20^\circ<\delta<+90^\circ$) \\
\cline{2-9}
\multicolumn{8}{c}{}\\[-2ex]
\cline{2-9}
Tracks  &  2 & 3.5 & 1.5 & 0.0 &  0 & 0.0 & 0.5 & 0.0 & NEAR HORIZONTAL \\
\cline{2-9}
Showers &  8 & 5.9 & 6.4 &  4.2   &  1 & 0.2 & 2.6 &  1.9   & 
($-20^\circ<\delta<+20^\circ$) \\
\cline{2-9}
\multicolumn{8}{c}{}\\[-2ex]
\cline{2-9}
Tracks  &  0 & 0.2 & 1.6 & 0.0 &  0 & 0.0 & 0.6 & 0.0 & DOWNGOING \\
\cline{2-9}
Showers & 11 & 0.6 & 6.5 &  8.0   &  3 & 0.0 & 2.9 &  3.5   & 
($-90^\circ<\delta<-20^\circ$) \\
\cline{2-9}
\multicolumn{10}{c}{}\\[-2.5ex]
 \multicolumn{1}{c}{}
&\multicolumn{4}{c}{30 -- 300~TeV} & \multicolumn{4}{c}{300 -- 3000~TeV}
\end{tabular}
\end{center}
\vspace{-0.5cm}
\caption{\it Data, atmospheric background, excess from a 
$E^{-2}$ diffuse flux, and excess from eikonal collisions 
of cosmogenic neutrinos ($M_5 = 1.7$~TeV, $m_c = 1$~GeV) in 988 days.
\label{tab:events2}}
\end{table}
It is apparent that the sum of the atmospheric background and
our hypothesis provides the most accurate fit of the data.
In particular, the likelihood ratio $\lambda$ \cite{Agashe:2014kda}
\beq
-2\ln \lambda = \sum_i^N 2 \left( E_i - X_i + X_i\,\ln {X_i\over E_i} 
\right)\,,
\eeq
where $E_i$ is the prediction, $X_i$ 
the data and $N$ the number of bins, 
gives a significant difference between both hypotheses:
\beq
-2\ln \lambda^{NP}= 5.9\;,\qquad -2\ln \lambda^{E^{-2}}= 15.4\;.
\eeq
If the 5 ambiguous tracks were included in the analysis, 
we would obtain similar values:
\beq
-2\ln \lambda^{NP}= 7.3\;,\qquad -2\ln \lambda^{E^{-2}}= 15.1\;.
\eeq

\section{Summary and discussion}
The observation by IceCube of 37  events with
energy above 30~TeV during the past 3 years is with no doubt
a very remarkable and interesting result. Their analysis has 
shown (and ours confirms) that atmospheric neutrinos are 
unable to explain the data. Therefore, IceCube has most
certainly discovered a neutrino flux of different origin. 
We think, however, that the determination of the 
nature and the possible origin of this flux is 
still work in progress.

The events observed do not exhibit a clear
preference for the galactic disc and/or
the galactic center. The best fit of the data by IceCube
has been obtained using a diffuse cosmic
flux with a spectrum proportional to $E^{-2}$. Since neutrinos can 
propagate without significant  energy losses 
from very distant sources, an isotropic diffuse flux generated 
by the ensemble of all extragalactic sources in the universe
is indeed expected.

This hypotesis, in principle, implies equipartition between
the 3 neutrino flavors and a given  distribution of 
zenith angles. Regarding the first point, 
it gives around 1 muon event per 4.5 showers, whereas 
the excess that we find in the data is around 18.6 showers 
and no muons (4 observed, 4.5 atmospheric events expected; if 
the muon background were included, we would 
observe 9 events but 12.9 expected). The expected number of
tracks could be smaller if the efficiency to
detect charged current $\nu_\mu$ interactions (the effective 
IceCube mass for these processes described in 
in \cite{Aartsen:2013jdh}) in IceCube were lower 
than assumed (see the discussion in \cite{Aartsen:2014muf}).
In any case, it seems clear that
the uncertainties and the low statistics still available
make the muon count compatible both with IceCube's $E^{-2}$ hypothesis  
and also with the basic result in \cite{Mena:2014sja} (that we
subscribe): muon topologies are well 
explained by the atmospheric flux, the only significant
excess appears in the number of showers.
As for the zenith angle distribution, we have distinguished 3 regions
of similar angular size: downgoing, near-horizontal and upgoing
directions. At PeV energies the Earth is (partially)
opaque only to upgoing
neutrinos ($+20^\circ<\delta<+90^\circ$). Therefore,  
IceCube's diffuse-flux hypothesis implies a similar number of events
in the downgoing and horizontal bins. The data, however,
reveals an excess of 13.4 shower events in the first bin but just 
2.9 from horizontal directions. Of course, the low statistics
gives little significance to these 
discrepancies\footnote{The main difference between IceCube's 
analysis and ours is that they seem to treat the atmospheric 
neutrino flux from the prompt decay of charmed hadrons as 
an error bar, whereas in our case it dominates over the
flux from $\pi$ and $K$ decays at  
$E>10^{5.5}$~GeV (see Fig.~1).}, but
it also leaves plenty of room for alternative explanations.

We have proposed a scenario where the IceCube excess appears only 
in showers (no muon topologies) from downgoing and near-horizontal 
directions (no upgoing events) in a 2:1 ratio. It seems to provide
a more accurate fit of the data than the $E^{-2}$ flux hypotesis.
The excess events are caused by {\it exotic} very-soft interactions of 
cosmogenic neutrinos, whose flux can be
estimated with some accuracy assuming that the $10^{10}$--$10^{11}$ 
cosmic rays observed by AUGER \cite{Settimo:2012zz} are 
protons\footnote{Notice also that these
soft interactions experienced by 
cosmogenic neutrinos are unconstrained by
AUGER \cite{Abreu:2013zbq}, since the energy deposited
in the atmosphere, of order
$yE\approx 10^6$~GeV, is below threshold.}. The much
larger energy of these neutrinos, around $10^9$~GeV, prevents
them from reaching IceCube from below, suppressing the flux
in a $\approx 50\%$ already from horizontal directions.

We have defined a TeV gravity model that provides a neutrino-nucleon
cross section with the precise features that are required. It
should be considered as a particular realization of the generic type of
models \cite{Dvali:2010jz,Dvali:2012mx} where UV physics is dominated by 
long-wavelength degrees of freedom. 
We think that an increased statistics at IceCube will establish
whether new physics (see also \cite{Esmaili:2013gha})
is necessary in order to interpret the data.

\section*{Acknowledgments}

We woud like to thank Carlos P\'erez de los Heros and 
Monica Verducci for useful discussions.
This work has
been supported by MICINN of Spain (FPA2010-16802, FPA2013-47836, and  
Consolider-Ingenio {\bf Multidark} CSD2009-00064), by Junta de 
Andaluc\'\i a (FQM101,3048,6552), 
and by MIUR of Italy under the program Futuro in Ricerca 2010
(RBFR10O36O).

\end{document}